# A quantitative framework for evaluating architectural patterns in ML systems

Simeon Emanuilov and Aleksandar Dimov

Department of Software Technologies, Faculty of Mathematics and Informatics
Sofia University "St. Kliment Ohridski"
5 James Bourchier Blvd., 1164 Sofia, Bulgaria
{ssemanuilo, aldi}@fmi.uni-sofia.bg

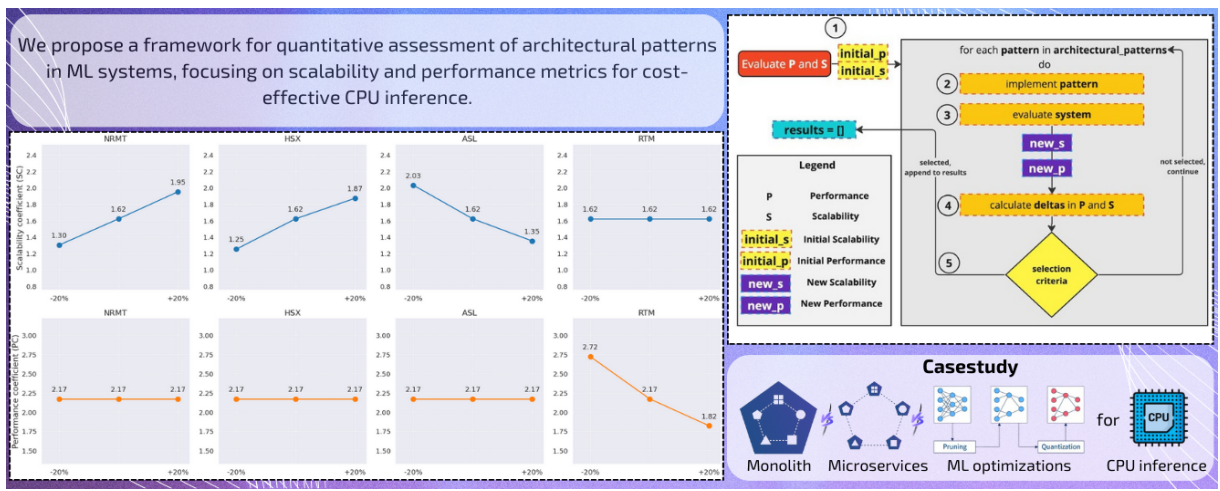

**Abstract:**
Contemporary intelligent systems incorporate software components, including machine learning components. As they grow in complexity and data volume such machine learning systems face unique quality challenges like scalability and performance. To overcome them, engineers may often use specific architectural patterns, however their impact on ML systems is difficult to quantify. The effect of software architecture on traditional systems is well studied, however more work is needed in the area of machine learning systems. This study proposes a framework for quantitative assessment of architectural patterns in ML systems, focusing on scalability and performance metrics for cost-effective CPU-based inference. We integrate these metrics into a systematic evaluation process for selection of architectural patterns and demonstrate its application through a case study. The approach shown in the paper should enable software architects to objectively analyze and select optimal patterns, addressing key challenges in ML system design.

# 1. Introduction

Machine learning (ML) is revolutionizing industries from healthcare to finance, introducing new paradigms in software system design [1]. However, ML-enabled systems present unique challenges for software architects and engineers, distinct from those of traditional systems [2] [3]. These challenges encompass computational complexity, deployment management, data dependencies, and monitoring [4] [5] [6].

Deep neural networks and other complex ML models often demand significant computational resources for both training and inference. Their deployment and management frequently require updates and versioning, which can complicate system consistency and maintainability [7]. In distributed architectures, additional challenges arise from data dependencies and communication overhead. Fundamentally, ML systems differ from traditional systems as they are not deterministic and have an ability to learn and adapt without extensive hard coding [2]. This contrasts with systems where engineers explicitly define input-output relationships and necessitates specialized design strategies and tactics.

Architectural patterns represent time-tested strategies for software design and offer promising solutions to these ML-specific challenges. While these patterns have been extensively studied in various contexts [8] [9] [10] [11], there's a growing need to adapt them specifically for ML systems [12] [13] [14].

In the production of ML systems, scalability and performance are crucial considerations [4] [7]. Scalability refers to a system's ability to efficiently handle increased data or user traffic, while performance focuses on processing and inference speed. Balancing these aspects is essential for effective production ML systems [15].

To address these challenges and optimize ML system design, the goal of this paper is to propose a framework for quantitatively assessing architectural patterns in ML systems. Our approach emphasizes scalability and performance metrics, particularly for CPU-based inference scenarios. This systematic evaluation process guides pattern selection to enhance these crucial aspects in ML deployments, prioritizing cost-effective solutions for resource-intensive models. Specifically, our framework evaluates patterns for ML systems deployed using REST API and gRPC strategies, aiming to optimize system architecture for efficient resource utilization.

The remainder of this paper is organized as follows: *Section 2* provides background on architectural patterns and ML system characteristics and the notations used in the paper. *Section 3* reviews related work. *Section 4* introduces our quantitative assessment framework, then *Section 5* proposes an evaluation process for assessing patterns' impact on ML system scalability and performance. *Section 6* demonstrates the application through experiments. *Section 7* concludes with key findings and future directions.

## 2. Background

### 2.1 Software architectural patterns

Software architects often apply proven solutions to recurring design challenges, enhancing system qualities like maintainability, durability and scalability [16]. These solutions are termed architectural styles, architectural patterns, or design patterns in literature. All these concepts solve recurring problems in the context of system design; however, the level of detail is different.

Some sources use them interchangeably [17], others differentiate them based on scope, application [18], or their role in the software development process [19]. Architectural patterns typically represent high-level design decisions and structural principles, while design patterns address more specific coding challenges.

Although strict differentiation between these concepts is important, it is not central to this paper's scope. Our assessment framework (*Section 4*) and patterns evaluation process (*Section 5*) should be applicable across all of them, and the focus is on quantitatively measuring the effects, when applied to ML systems [20] [21].

Therefore, throughout this work, we use the term "architectural patterns" to encompass all these related concepts. The proposed framework provides a systematic approach to evaluate the impact of these patterns on ML system scalability and performance.

### 2.2 Traditional versus ML-based software systems

In our scope, traditional software systems are primarily deterministic, characterized by predefined rules and structured data processing pathways [2]. Developers explicitly program the logic, resulting in predictable outputs for given inputs. These systems are typically developed using a top-down approach and offer consistent performance within predefined operational parameters.

In contrast, ML-based systems have the capacity to learn patterns and make decisions from data without explicit programming of rules (*Fig. 1*). This adaptive capability is a hallmark of Artificial Intelligence (AI), with Machine Learning being a subset of AI, and Deep Learning (DL) a further specialized subset of ML.

This work uses the term "ML-based systems" broadly, focusing specifically on computationally intensive systems that often, but not exclusively, employ DL techniques. These systems incorporate complex models, particularly those using neural network architectures, to perform tasks that traditionally require explicit algorithmic solutions. While parts of the machine learning community emphasize ML as primarily a tool rather than a system, the integration of ML components into larger software ecosystems creates unique architectural challenges. Our focus is on these ML-enabled systems, where ML models and algorithms are integral parts of a broader software architecture, requiring specialized design considerations.

Typical approaches in ML systems include deep neural networks, convolutional neural networks, and recurrent neural networks. They may also utilize other computationally demanding methods such as large-scale random forests or complex ensemble models. The common threads among such approaches are their computational intensiveness and ability to adapt and improve performance over time as they are exposed to more data.

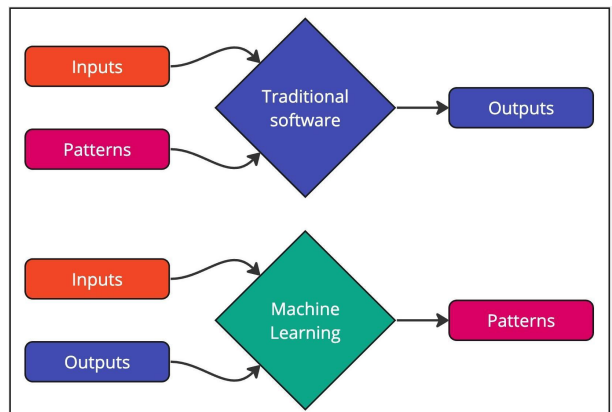

*Fig 1 Comparison of traditional software development and machine learning processes. Traditional systems have explicitly defined logic, while ML systems learn input-output mappings from data, adapting over time* [2]

*Table 1* highlights key differences between traditional and ML-based systems, underscoring the need for specialized approaches in architecting such systems. The criteria used in the comparison are based on a comprehensive literature review (*Section 3*) and practical use cases [22] [23].

*Table 1 Comparison of traditional and ML-based software systems*

| Criteria | Traditional | ML-based |
| --- | --- | --- |
| Operation consistency | Uniform operation | Improves with usage |
| Rule definition | Predetermined logic | Data-driven logic |
| System quality | Depends on code quality | Depends on code, data, and hyperparameters |
| Objectives | Functional and non-functional requirements | Additional optimization metrics (e.g., accuracy, precision/recall) |
| Testing | Simplified procedures | Complex experiments with baselines and ground truths |
| Computational requirements | Generally predictable | Often intensive and variable |

The distinctions outlined in *Table 1*, introduce unique challenges in the design, implementation, and optimization of ML-based systems [2] [4] [7] [15]. The following section will explore them in detail, setting the stage for our discussion on architectural patterns for ML-based systems.

## 2.3 Challenges in production of ML-based systems

The unique challenges of ML systems stem from their inherent characteristics and require specialized approaches [4] [24]. When applying architectural patterns to such systems, it's important to quantitatively measure the impact of patterns on system performance and scalability, to address the challenges. Measurement should allow for objective evaluation of pattern effectiveness in the context of ML-specific requirements and constraints [21] [25]. Most significant challenges in ML-based software systems production are listed below.

A) Computational complexity

The computational complexity of ML models, especially deep neural networks, is a primary challenge [4] [7] [26]. These models are often resource-intensive, requiring significant computational power during inference. Executing these models on CPUs can be resource-consuming, necessitating optimized system architecture for efficient execution.

B) Deployment management

Managing and deploying ML models within the system architecture presents another challenge [5] [26]. ML models frequently need updates and versioning to incorporate new data and improve performance. Ensuring consistency and compatibility across system components can be complex, particularly with evolving models. Efficient resource allocation and scaling of ML model components is important, as resource requirements may vary based on specific models and workloads.

C) Data dependencies

Data dependencies and communication overhead pose significant challenges [6]. ML models heavily rely on data, making efficient data exchange between system components strategic [26] for overall performance. When distributing ML components across multiple services or nodes, managing data dependencies and minimizing communication overhead become a key consideration.

D) Monitoring

Monitoring performance and detecting anomalies is more challenging than in traditional systems, as there is additional complexity in identification and diagnosis [25]. Robust monitoring and error handling mechanisms are essential for ensuring system reliability and performance, especially in distributed architectures.

All these challenges underscore the need for careful consideration when selecting and applying architectural patterns to ML-based systems. With a multitude of patterns available, choosing the most appropriate ones becomes fundamental.

Our research aims to provide a comprehensive framework for evaluating the impact of various architectural patterns on ML-based systems, helping architects identify which patterns best address their unique characteristics. This approach not only accounts for the challenges and their potential implications but also guides the selection process from the wide array of available options, ensuring optimal system design.

# 3. Related work

Several research directions are directly related to our work – (1) general frameworks for pattern evaluation; (2) Architectural pattern selection and classification; (3) Elaboration and research on patterns specifically targeted at ML systems and (4) Performance and scalability of ML Systems. Although systematic literature survey on these topics is beyond the scope of this paper, we applied the following strategy to explore the related work:

1. We utilized mainly ResearchGate, IEEE Xplore® and Google Scholar, focusing on peer-reviewed publications.
2. We focused on both general pattern evaluation frameworks and ML-specific research, as this is the main contribution of our work.
3. For general frameworks, we used search terms such as "architectural pattern evaluation", "design pattern assessment frameworks" and "software pattern quality metrics".
4. For ML-specific research, our search terms included "machine learning architectural patterns", "ML system design patterns", "scalability in ML systems" and "performance optimization for ML".

This strategy yielded over 190 initial results across various domains. After screening for relevance and quality, we thoroughly reviewed approximately 60 sources, with 46 directly cited in this study. They include not only academic papers, but also technical blogs, books, and official documentation, providing a diverse range of perspectives. The selected sources, spanning from 1990 to 2024, offer a comprehensive view of both foundational and cutting-edge research, forming a solid foundation for our work.

## 3.1 General frameworks for pattern evaluation

Several works have proposed frameworks for evaluating design patterns in different contexts. The Software Architecture Analysis Method (SAAM) [27] is one of the first systematic methods for analyzing software architectures. While primarily focused on modifiability, it can be used for various quality attributes. SAAM involves developing scenarios, describing the architecture, classifying and evaluating scenarios, and overall evaluation.

The Architecture Tradeoff Analysis Method (ATAM) [11] builds on SAAM, assessing how well an architecture satisfies multiple quality goals and their trade-offs. It consists of nine steps across four phases, involving a wide range of stakeholders and utilizing various analysis techniques.

Briand et al. [28] introduced COMPARE (COMPrehensive Architecture Evaluation), integrating scenario-based analysis, qualitative pattern-based evaluation, and quantitative metrics assessment. COMPARE aims to provide a consistent framework regardless of architecture representations and granularity, proposing an iterative process for architecture evaluation and refinement.

The Cost Benefit Analysis Method (CBAM) [29] provides a structured approach for evaluating architectural design decisions based on economic considerations. CBAM builds on

ATAM by quantifying the costs, benefits, and uncertainty of architectural strategies using scenarios, utility-response curves, and stakeholder input.

Later, Babar and Gorton [30] conducted a comparative study of scenario-based software architecture evaluation methods, including SAAM and ATAM and proposed a framework for comparing such methods based on context, stakeholders, structure, and reliability.

Kassab et al. [31] developed a quantitative approach to assess the impact of architectural patterns on performance and security. They used matrix transformations to calculate how patterns support these quality attributes. Sant'Anna et al. [32] compared aspect-oriented and object-oriented implementations of design patterns, evaluating separation of concerns, coupling, cohesion, and size metrics in relation to maintainability and reusability.

Al-Obeidallah [33] presented a metrics-based approach to assess design patterns' impact on software maintainability and understandability, using size, coupling, and inheritance metrics. Xi et al. [34] proposed a service pattern assessment framework using a Quantitative Service Pattern Description Language (SPDL-Q), enabling the analysis of service patterns from multiple perspectives.

Our approach builds on these existing frameworks by providing a quantitative assessment specifically focused on architectural patterns in ML systems. We incorporate analysis like SAAM and ATAM, while utilizing metrics to measure impact on quality attributes like Al-Obeidallah's work. However, we adapt our framework to the unique challenges of ML architectures and consider quality attributes, particularly important for machine learning systems. In *Section 3.5* we present more details about our contribution and direct comparison with ATAM in *Table 2*.

## 3.2 Pattern selection and classification

Research in pattern selection and classification has yielded frameworks applicable to various domains. Hussain et al. [35] proposed a framework for automated classification and selection of software design patterns using text categorization and unsupervised learning techniques. Their approach, while focused on software design patterns, provides insights that could be extended to ML systems. Petter et al. [36] offered an evaluation framework for patterns in design science research, proposing criteria such as plausibility, effectiveness, and reliability. These criteria provide a useful starting point, though our work extends this by providing concrete metrics and evaluation processes for quantitative assessment in ML systems.

## 3.3 ML-specific patterns

Recent research has begun to focus on patterns specifically for AI systems. Washizaki et al. (2019) [13] explored software engineering patterns tailored for those systems, emphasizing the need for systematic collection and classification. Their study highlights the growing complexity of ML systems and the necessity for specialized patterns.

Lakshmanan, Robinson, and Munn (2020) [20] provided a guide for data scientists, detailing patterns across the ML process and addressing common challenges in ML system optimization. Later, Heiland et al. (2023) [37] conducted a comprehensive literature review to gather and categorize design patterns for AI-based systems, reflecting the evolving nature of design patterns in this context.

## 3.4 Performance and scalability in ML systems

Several studies have addressed performance and scalability issues in ML systems. Yan et al. (2015) [38] developed performance models for distributed deep learning systems, contributing to our understanding of scalability optimization in ML contexts. Boden et al. (2017) [39] benchmarked data flow systems for scalable ML, highlighting challenges in handling high-dimensional data, a critical aspect of large-scale ML algorithms. Olston et al. (2017) [40] introduced TensorFlow-Serving, focusing on optimizing model lookup and inference paths, which is fundamental for scalability and efficiency in ML systems. Tian et al. (2023) [41] presented a scalable ML framework for large-scale systems, offering insights into the applicability of ML in complex environments.

## 3.5 Comparative analysis

Among the various architectural analysis methods discussed in the literature, ATAM stands out as a particularly influential approach, incorporating principles from SAAM.

Our method builds upon them, extending and adapting it for the specific needs of ML systems. This study addresses the need for a rigorous, quantitative evaluation of design patterns in ML systems [13], proposing a structured evaluation process that focuses on scalability and performance.

To illustrate the key similarities and differences between ATAM and our proposed method, we present a comparison in *Table 2*. It shows how our approach adapts architectural analysis to the unique specificities of ML systems, maintaining the core principles of systematic architectural evaluation.

*Table 2 Key differences between ATAM and your method*

| Aspect | ATAM | Our method |
|---|---|---|
| Focus | General software systems | ML-specific systems |
| Quality attributes | Generic | ML-centric (e.g., scalability, performance for CPU-based inference) |
| Analysis approach | Qualitative | Quantitative and qualitative |
| Metrics | Limited, mostly qualitative | Specific, quantifiable metrics (e.g., NRMT, HSX, SC) |
| Tradeoff analysis | General | ML-specific tradeoffs |
| Iteration model | Spiral | Spiral with ML-specific refinement |
| Stakeholder involvement | High | High, with ML expertise emphasis |
| Scenario development | Generic | ML-focused scenarios |

Our work also integrates insights from other general pattern evaluation frameworks, ML-specific pattern research, and performance optimization studies, mentioned in this section.

By doing so, we aim to fill a crucial gap in the literature and provide practical tools for ML system architects and developers.

## 4. Quantitative assessment framework

This section introduces the framework for quantitative assessment of architectural patterns in ML systems, focusing on scalability and performance metrics. These metrics provide an objective basis for evaluating the impact of architectural patterns and guiding their selection. They were selected based on their ability to address the challenges described in *Section 2.3*. They provide a comprehensive view of ML system scalability and performance, as evidenced by their prevalence in existing literature [3] [21] [25] [42] [43] [44].

The framework includes three scalability-related metrics:
- *Number of Requests at Max Throughput (NRMT)*.
- *Horizontal Scaling Index (HSX)*.
- *Scalability Coefficient (SC)*.

And three performance-related metrics:
- *Average System Load (ASL)*.
- *Response Time Median (RTM)*.
- *Performance Coefficient (PC)*.

Each metric is described using the following structure: (1) metric meaning, (2) calculation process, (3) intuition behind the metric's applicability.

The framework key outputs are the *Scalability Coefficient (SC)* and *Performance Coefficient (PC)*. These are high-level metrics that provide a synthesized measure of a system's scalability and overall performance, allowing for effective comparison between different patterns.

For model deployment strategy, this study focuses on the REST API/gRPC approach, commonly used in various ML deployment scenarios [2] [22]. These protocols enable efficient communication between clients and ML services, allowing for seamless integration of ML models into distributed systems and web applications. REST API provides a standardized, resource-oriented architecture, while gRPC offers high-performance, language-agnostic remote procedure calls, both facilitating scalable and flexible ML model serving. It's important to also note that our work specifically targets computationally intensive ML systems, often employing DL techniques, as defined in *Section 2.2*.

### 4.1 Scalability metrics

### 4.1.1 Number of Requests at Max Throughput

The *Number of Requests at Max Throughput (NRMT)* metric measures the maximum number of simultaneous requests a system can handle before deviating from linear scalability.

Linear scalability is an ideal scenario where system responsiveness scales directly proportional to the number of requests. In practice, systems often deviate from this as requests increase, due to factors such as increased contention for CPU, memory, and network resources.

**Calculation process:**
- Establish a controlled environment.
- Measure system performance using load testing tools like JMeter[1], BlazeMeter[2], or a custom script (see *Appendix - Listing 2*).
- Gradually increase simultaneous requests and measure performance after each increment.
- Identify the point where performance deviates from linear scalability; this number of requests is the *NRMT*.

**Intuition:** A higher *NRMT* indicates that the system can handle more concurrent requests before performance degradation. This suggests better scalability and efficient utilization of resources under increased load. It's a key indicator of a system's ability to maintain performance as demand grows.

### 4.1.2 Horizontal Scaling Index

The *Horizontal Scaling Index (HSX)* reflects a system's horizontal scalability, considering: (1) time to scale, (2) ease of scaling, (3) costs of scaling the system.

A function of these metrics constitutes the basic coefficient.

**Calculation:**
$$HSX = (HSX_t * HSX_e * HSX_c)^{-1}$$

**where:**
- $HSX_t$ - time to scale horizontally (minutes to add new instances) - represents the time to scale horizontally, measured in minutes to add new instances. It quantifies the speed of integrating new instances into a system to double the *NRMT*. This metric focuses on the duration from initiating the scaling process to the moment the system achieves the expanded capacity. Shorter durations are preferable, reflecting a system's agility in adapting to escalating demands.
- $HSX_e$ - ease of horizontal scaling, $HSX_e \in \{1, 2, 3\}$ - how straightforward it is to add new instances. It assesses how straightforward it is to add new instances, considering factors such as manual intervention required, process complexity, and potential for errors or complications. Lower scores indicate easier scaling:
  - $HSX_e = 1$ (easy scaling).
    **Example:** An ML system deployed on a cloud platform with fully automated scaling. The system automatically adds new instances when traffic increases without manual intervention, using services like AWS Auto Scaling[3].

---

[1] Apache JMeter™, https://jmeter.apache.org/
[2] BlazeMeter Continuous Testing, https://www.blazemeter.com/
[3] AWS Auto Scaling, https://aws.amazon.com/autoscaling/

- ○ $HSX_e = 2$ (moderate scaling).
  **Example:** A setup requiring some manual steps to add new instances, but with partial automation. This might involve manually initiating a script that automates the rest of the scaling process.
- ○ $HSX_e = 3$ (difficult scaling).
  **Example:** A scenario where adding new instances is entirely manual and time-consuming. This could involve manually configuring and deploying each new instance, including server setup, software installation, network configuration, and system integration.
- $HSX_c$ - represents the cost of doubling *NRMT* with new instances, measured in thousand EUR per month. It reflects the financial implications of horizontal scaling, including direct costs of adding new nodes to double the system's max throughput.

**Intuition:** While $HSX_e$ is more subjective, a higher overall *HSX* indicates better system scalability, reflecting faster server addition, lower scaling costs, and smoother operations. Although the component metrics measure diverse aspects (time, ease, cost), their combination provides a holistic view of a system's horizontal scalability, enabling effective comparison and evaluation of different systems.

### 4.1.3 Scalability Coefficient

The *Scalability Coefficient (SC)* is a comprehensive measure of a system's scalability. It combines several key metrics: the *Number of Requests at Max Throughput (NRMT)*, the *Average System Load (ASL)*, and the *Horizontal Scaling Index (HSX)*. The *SC* is calculated as follows:

**Calculation:**
$$SC = \frac{NRMT * HSX}{ASL}$$

**Intuition:** A higher *SC* indicates a system that can scale easily and efficiently. It suggests that the system can handle a high number of requests, is relatively easy to scale horizontally, and maintains a manageable system load while doing so. This metric is particularly valuable for comparing the scalability of different systems or architectural patterns in ML contexts, where the ability to handle increasing computational demands is important.

## 4.2 Performance metrics

### 4.2.1 Average System Load

*Average System Load (ASL)* measures the system's resource utilization (CPU, memory, I/O) when handling the maximum number of requests at its peak throughput (*NRMT*). It represents the average load when the system processes the highest number of concurrent requests without deviating from linear scalability.

The *ASL* can be calculated using the Unix uptime command[4] in a maximally isolated environment, averaging load information from the preceding 1, 5, and 15-minute intervals.

**Calculation:**

$$ASL = \frac{1}{M} \sum_{j=1}^{M} load_j$$

**where:**
- *load$_j$* - the number of runnable processes over the preceding 1, 5, and 15-minute intervals.
- *M* - the number of intervals measured.

**Intuition:** A lower *ASL* indicates an optimized system where operations consume fewer resources. It suggests efficient utilization of CPU, memory, and I/O, with the system not overburdened at peak workload. The *ASL* metric allows for comparison between different systems or tracking performance changes in a single system over time.

### 4.2.2 Response Time Median

The *Response Time Median (RTM)* represents the middle value in a sorted list of response times for a set of consecutive requests. This metric is less sensitive to outliers than the average, providing a measure of system performance.

To determine the *RTM*, first sort all response times in ascending order. For an odd number of response times, the *RTM* is the middle value. For an even number, it's the average of the two middle values.

**Calculations:**

$$RTM = \begin{cases} RT_{\left(\frac{N_r + 1}{2}\right)}, & if\ N_r\ is\ odd \\ \dfrac{RT_{\left(\frac{N_r}{2}\right)} + RT_{\left(\frac{N_r}{2}+1\right)}}{2}, & if\ N_r\ is\ even \end{cases}$$

**where:**
- $RT_i$ - represents the i$^{\text{-th}}$ smallest response time in the sorted list.
- $N_r$ - represents the number of requests (typically 1000 or more).

**Intuition:** A lower *RTM* indicates better system performance, as it means most requests are processed quickly. The median provides a robust measure of central tendency, less affected by extreme values compared to the mean. This makes *RTM* particularly useful for assessing performance in ML systems where occasional outliers (*e.g.*, due to complex computations) might skew an average-based metric.

---
[4] Uptime manual page, https://www.unix.com/man-page/osx/1/uptime/

### 4.2.3 Performance Coefficient (PC)

The *Performance Coefficient (PC)* is a metric that reflects the overall speed and efficiency of the system. It provides a basis for comparing different system configurations. The *PC* is calculated as follows:

**Calculation:**

$$PC = \frac{N_r}{RTM}$$

**where:**
- $N_r$ - the number of requests used to calculate the *Response Time Median (RTM)*.
- *RTM* - the *Response Time Median*.

It's important to use the same $N_r$ value across different experiments to maintain consistency in the scale of computations.

**Intuition:** A higher *PC* indicates an optimized system with better performance and faster response times.

The *Scalability Coefficient (SC)* and *Performance Coefficient (PC)* are primary metrics derived from lower-level metrics such as *NRMT*, *HSX*, *ASL*, and *RTM*. These secondary metrics capture specific aspects of system behavior under varying loads, contributing to the calculation of the primary metrics. *Fig. 4* provides a sensitivity analysis and can contribute to the understanding of the rationale behind them.

By applying this framework, software engineers can systematically assess and compare the impact of different architectural patterns on the scalability and performance of ML systems. It provides a quantitative basis for decision-making in system design and optimization. However, it's important to note that there are some limitations and additional considerations to this framework, which are discussed in more detail in *Section 6.4*.

The evaluation process proposed in the next section builds upon this framework, offering a step-by-step process for evaluating and selecting architectural patterns tailored to the unique demands of ML systems.

## 5. Evaluation process for architectural patterns

This section outlines the process for evaluating architectural patterns in ML systems. The output is a "result list" of patterns that meet specific selection criteria based on their impact on scalability and performance. The evaluation process involves comparing a baseline ML system (*System A*) to systems with architectural patterns applied (e.g., *System B*), using the quantitative metrics outlined in *Section 4*.

The primary focus is on scalability and performance, though this approach can be adapted to evaluate other qualities such as reliability and maintainability. By applying patterns to a baseline system and measuring changes in key metrics, this approach enables quantitative assessment of their impact, providing insights for system architects.

The evaluation process consists of the following steps (illustrated in *Fig. 2*):

1. Evaluate the initial scalability and performance of the baseline ML system (*System A*) using metrics from *Section 4* (*NRMT*, *HSX*, *ASL*, *RTM*, *SC*, and *PC*).
2. Implement a target design pattern on the system.
3. Re-evaluate the system's scalability and performance.
4. Calculate the deltas in scalability ($delta\_s$) and performance ($delta\_p$) by subtracting the initial *SC* and *PC* values from their post-implementation values, i.e., $delta\_s = new\_scalability - initial\_scalability$ and $delta\_p = new\_performance - initial\_performance$.
5. If there's a positive improvement in either quality attribute (i.e., $delta\_p * delta\_s >= 0$ and $delta\_p + delta\_s > 0$), the pattern meets the selection criteria and is added to the result list. If not, the evaluation moves to the next one.

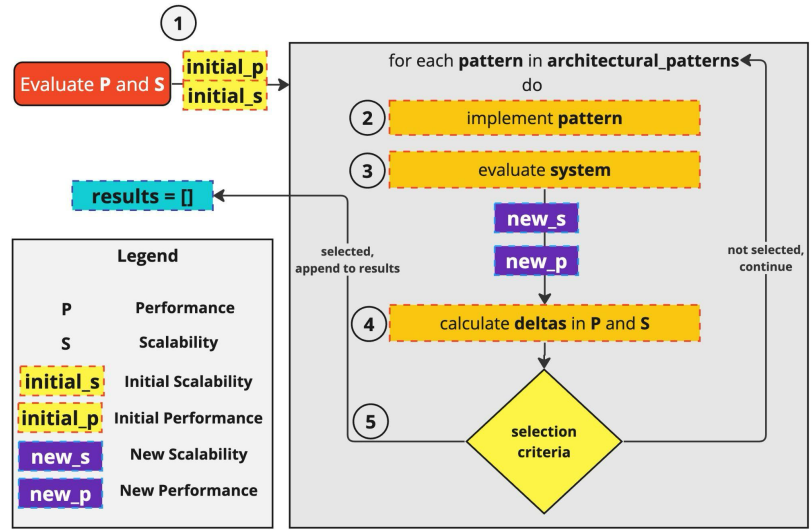

**Fig. 2** *A diagram with the proposed evaluation process*

A pseudo code with additional comments on each step is shown in *Listing 1*.

```
# Step 1: Evaluate the initial scalability and performance on a given baseline ML system (MLS)
initial_scalability, initial_performance = evaluate(mls)

# Keep the promising patterns in a list
result = []

for pattern in architectural_patterns:
    # Step 2: Implement the current architectural pattern.
```

```
new_mls = implement_pattern(pattern)

# Step 3: Re-evaluate the scalability and performance of the system
new_scalability, new_performance = evaluate(new_mls)

# Step 4:Calculate the deltas in scalability and performance.
delta_p = new_performance - initial_performance
delta_s = new_scalability - initial_scalability

# Step 5: If there is a positive improvement in one of the
# quality attributes,
# add the pattern to the result list.
# We need to make sure that there are no deltas,
# pointing in opposite directions (for example: -3 and 2)
# and also check if the final result is positive.
if (delta_p * delta_s) >= 0 and (delta_p + delta_s) > 0:
    result.append(pattern)
```

*Listing 1* A pseudo-code used for evaluating target patterns

**Where:**
- *architectural_patterns* - a list of input software architectural patterns/styles for evaluation.
- *evaluate()* - a function that calculates and returns scalability and performance coefficients
- *MLS* - a baseline Machine Learning system (System A).
- *implement_pattern()* - a function that modifies the system to incorporate a new architectural pattern.

This evaluation process provides a systematic approach to evaluating architectural patterns for ML systems, extending previous work on architecture evaluation methods (*Section 3.1*). However, it's important to understand how to interpret and apply the results effectively. The core of our evaluation lies in the changes to the *Scalability Coefficient (SC)* and *Performance Coefficient (PC)*. These primary metrics capture complex system behaviors in a simplified form. For instance:

1. A decrease in *Response Time Median (RTM)* will increase the Performance *Coefficient (PC)*.
2. An increase in the *Number of Requests at Max Throughput (NRMT)* will increase the *Scalability Coefficient (SC)*.

It is worth mentioning that architectural patterns often present trade-offs (*Section 6.4*). A pattern might enhance scalability while slightly degrading performance, or vice versa.

For example, a pattern that improves *HSX* but increases *RTM* would enhance *SC* but potentially reduce *PC*. To address this complexity, our method calculates *SC* and *PC* separately before combining them in the final evaluation step. This approach, illustrated in *Fig. 2*, allows us to:

1. Capture improvements in either scalability or performance.
2. Ensure that significant degradation in one aspect isn't masked by improvements in another.
3. Provide a balanced view of the pattern's overall impact on the system.

By using this nuanced approach, we can identify patterns that offer the best overall improvement to the ML system, even if they don't optimize every single metric. This method provides interested stakeholders with a tool for making informed decisions, accounting for the complex interplay between scalability and performance.

## 6. Case-study

This section presents a case study, by applying the evaluation process described in the previous section. We evaluate the following three systems:
1. *System A*: A monolithic ML application, described in *Section 6A*.
2. *System B*: System A rebuilt using a Microservice architectural pattern.
3. *System C*: System A with localized, ML-specific optimization techniques applied.

A) Baseline system characteristics (System A)

The evaluation uses a scalable API built with FastAPI[5] and TensorFlow serving[6], hosting multiple ML models:
- Classification models using Keras[7] with Xception[8] architecture.
- U-Net models for image background removal.
- VGG16 models for facial embedding extraction.

In addition to the inference system, a training pipeline is implemented using Apache Airflow[9]. This pipeline orchestrates periodic model retraining tasks, incorporating new data to improve model performance over time. The training process is designed to run on separate, compute resources to avoid interfering with the main inference system.

The system employs Redis[10] as a message broker for inter-service communication and PostgreSQL as the database. It's containerized using Docker and Docker Compose, ensuring consistency across development and deployment environments. The API, model services, and training pipeline use Python 3.11, with Celery[11] for distributed task processing.

Importantly, our evaluation focuses on CPU-based inference. While GPU acceleration is common in ML systems, our metrics in *Section 4* are tailored for CPU-based systems, emphasizing cost-effectiveness in scenarios where GPU resources may be limited or cost-prohibitive.

B) Other considerations

While microservices are known to improve scalability in traditional systems, the unique challenges of ML-based systems (see *Section 2.3*), can affect this architectural pattern's effectiveness. Our evaluation, therefore, specifically examines the microservice

---

[5] FastAPI framework, https://fastapi.tiangolo.com/
[6] TensorFlow serving repository, https://github.com/tensorflow/serving
[7] Keras, Deep Learning framework, https://keras.io/
[8] XCeption model and depthwise separable convolutions review, https://maelfabien.github.io/deeplearning/xception/
[9] Apache Airflow platform, https://airflow.apache.org/
[10] Redis, Real-time data platform, https://redis.io/
[11] Celery, Distributed task queue, https://docs.celeryq.dev/en/stable/index.html

pattern's impact on ML-based systems, considering their distinct characteristics and requirements.

This evaluation aims to provide insights into how different architectural patterns and optimization techniques affect the scalability and performance of ML systems in real-world scenarios, with a focus on CPU-based, cost-effective deployments.

## 6.1 High-level structural patterns

The baseline system consists of 3 servers, each with the following specifications: Intel Xeon-E 2136 (6c/12t, 3.3 GHz/4.5 GHz), 32 GB DDR4 ECC RAM, and 500 GB SSD NVMe. *Table 3* presents the results of applying our quantitative assessment framework (*Section 4*) to both the monolithic (*System A*) and microservice (*System B*) architectures. More detailed calculation instructions are provided in the *Appendix*.

For the microservice-based architecture (*System B*), the monolithic FastAPI application (*System A*) was decomposed into six microservices:
1. User management: Handles user authentication and authorization.
2. Data ingestion: Manages the ingestion and preprocessing of data for model training and inference.
3. Model training: Handles the training and validation of ML models.
4. Model serving: Deploys trained models and handles inference requests.
5. Job scheduling: Manages the scheduling and execution of periodic tasks, such as data ingestion and model retraining.
6. Monitoring service and database: Collects and saves system metrics and logs for performance monitoring and issue detection.

These microservices were implemented using a combination of FastAPI, TensorFlow Serving, and Celery (for asynchronous task processing). Inter-service communication was facilitated through RESTful APIs and message queues using Redis, similarly to *System A*.

*Table 3 Impact of high-level structural patterns on ML system scalability and performance. $N_r = 1000$ for both cases*

| Metric | System A | System B |
|---|---:|---:|
| *NMRT* | 256 | 256 |
| *HSX* | 0.04 | 0.16 ↑ |
| $HSX_t$ | 60 | 40 ↓ |
| $HSX_e$ | 2 | 1 ↓ |
| $HSX_c$ | 0.20 | 0.15 ↓ |
| *ASL* | 6.57 | 6.12 ↓ |
| *RTM* | 460 | 460 |
| **SC** | **1.62** | **6.97 ↑** |
| **PC** | **2.17** | **2.17** |

The following analysis interprets the quantitative results:

A) Scalability improvement

The *Scalability Coefficient (SC)* showed improvement, primarily due to changes in the *Horizontal Scaling Index (HSX)* and *Average System Load (ASL)*. This indicates that *System B*, employing a microservice architecture, is better equipped to handle increased loads and scale horizontally. The lower $HSX_e$ and $HSX_c$ values in *System B* suggest that adding new instances to handle increased traffic is more straightforward and cost-effective in the microservice architecture. This aligns with the general understanding of microservices' benefits but quantifies the improvement specifically for ML systems. It also gives a more numerical perspective of the benefits.

B) Resource utilization and performance

The lower *ASL* in *System B* implies more efficient resource utilization, likely due to the distributed nature of microservices and the ability to scale individual components independently. This is particularly beneficial for ML systems where different models or services may have varying resource requirements. The *Performance Coefficient (PC)* remained unchanged between the two systems. This suggests that while microservices improve scalability, they don't necessarily impact the base performance of the ML models themselves.

C) Architectural implications

These findings underscore the importance of evaluating high-level structural patterns for such systems. The microservice architecture demonstrates potential benefits for ML systems that prioritize scalability, offering more flexible and cost-effective horizontal scaling compared to a monolithic architecture (*System A*). However, it's important to note that microservice architectures often come with increased complexity in deployment and management. This trade-off should be considered alongside the scalability benefits when making architectural decisions for ML systems.

These results provide insights for ML system architects, particularly when designing systems that need to handle growing workloads and increasing user demands. The quantitative approach allows for more informed decision-making, balancing the benefits of improved scalability against potential increases in system complexity.

## 6.2 Localized, ML-specific optimization techniques

This part of the evaluation assesses the impact of localized, ML-specific optimization techniques on system scalability and performance. While the experiment described in Section 6.1., focused on high-level structural patterns (i.e. microservices), this one examines more targeted optimizations specific to ML workloads.

We applied two optimization techniques to the baseline system (*System A*) and compared the results (*Table 4*) with the new, optimized system (*System C*):
1. Weight: A technique that reduces neural network complexity by removing less important weights [45].

2. OpenVino model optimizer[12]: A tool that optimizes neural network models for inference on Intel hardware.

*Table 4* *The results of applying our evaluation process to assess the impact of these ML-specific optimizations*

| Metric | System A | System C |
|---|---|---|
| *NMRT* | 256 | 272 ↑ |
| *HSX* | 0.04 | 0.04 |
| *HSX$_t$* | 60 | 60 |
| *HSX$_e$* | 2 | 2 |
| *HSX$_c$* | 0.20 | 0.20 |
| *ASL* | 6.57 | 5.85 ↓ |
| *RTM* | 460 | 350 ↓ |
| **SC** | **1.62** | **1.94 ↑** |
| **PC** | **2.17** | **2.86 ↑** |

The analysis below examines the effects and considerations:

A) Performance and scalability improvements

*System C* shows improvements in both the *Performance Coefficient (PC)* and *Scalability Coefficient (SC)* compared to the baseline *System A*. These enhancements are primarily due to changes in the *Number of Requests at Max Throughput (NMRT), Average System Load (ASL)*, and *Response Time Median (RTM)* metrics. The optimizations have positively impacted the system's ability to handle increased loads and deliver faster response times.

B) Technique selection rationale

The choice of weight pruning and OpenVino model optimizer was based on several factors:
1. Relevance to system characteristics: Our ML system (described in *Section 6A*) uses deep learning models on Intel hardware, making these techniques particularly applicable.
2. Complementary benefits: Weight pruning addresses model complexity, while OpenVino optimizes for specific hardware, potentially offering synergistic improvements.
3. Industry prevalence: Both techniques are widely used in production ML systems, especially when deploying on resource-constrained devices or prioritizing inference speed [45].

However, it's important to note that other optimization techniques could have been considered, such as quantization, knowledge distillation, or hardware-specific optimizations for non-Intel platforms [46]. Future work could involve comparing the impact of these alternative techniques.

---

[12] OpenVINO™ model optimization guide, https://docs.openvino.ai/2024/index.html

C) Limitations and trade-offs

While both techniques showed positive impacts, it's important to consider potential trade-offs:
1. Weight pruning may slightly reduce model accuracy, which isn't captured in our current metrics, as it is more in the field of model evaluation.
2. OpenVino optimizations are hardware-specific, potentially limiting portability.

These trade-offs underscore the importance of comprehensive evaluation in real-world scenarios before full adoption.

D) Evaluation process versatility

*Fig. 3* provides a comparison of the *Scalability Coefficient (SC)* and *Performance Coefficient (PC)* for *Systems A, B,* and *C*, summarizing the impact of high-level structural patterns and ML-specific optimizations on ML system scalability and performance.

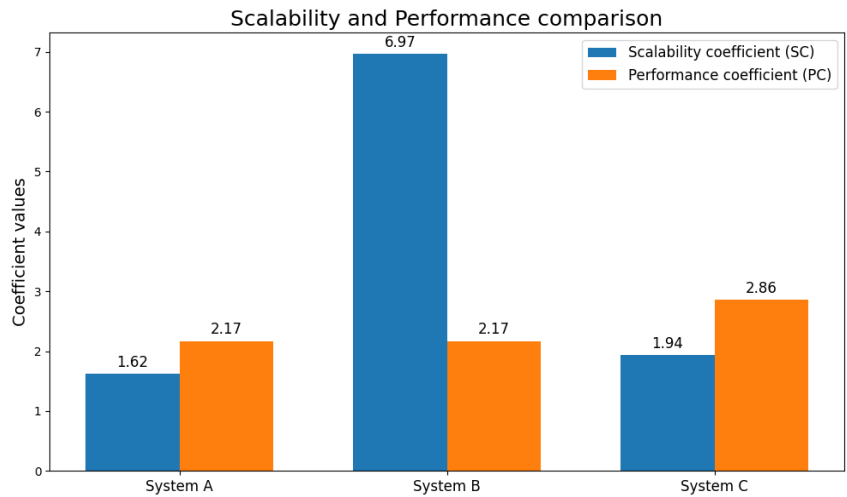

**Fig. 3** *Scalability and Performance comparison between Systems A, B and C*

The chart illustrates the improvement in scalability achieved by *System B* (microservice architecture) compared to the monolithic *System A*. It also shows the performance gains obtained through ML-specific optimizations in *System C*, demonstrating the cumulative effect of architectural patterns and targeted optimizations on ML system performance. Combination from *System B* and *C* could include the benefits from both experiments.

To further understand the factors influencing scalability and performance in ML systems, we conducted a sensitivity analysis using *System A* as the baseline. This analysis, presented in *Fig. 4*, examines how changes in intrinsic values (*NRMT*, *HSX*, *ASL*, *RTM*) affect the *Scalability Coefficient (SC)* and *Performance Coefficient (PC)*.

## 6.3 Sensitivity analysis

*Fig. 4* presents a sensitivity analysis of the *Scalability Coefficient (SC)* and *Performance Coefficient (PC)* to changes in intrinsic values (*NRMT*, *HSX*, *ASL*, *RTM*) for

*System A*. This analysis helps identify which intrinsic values have the most significant impact on the overall scalability and performance of the ML system, using *System A* as the baseline. The 20% interval purpose is to show the slope and direction of change of each individual metric.

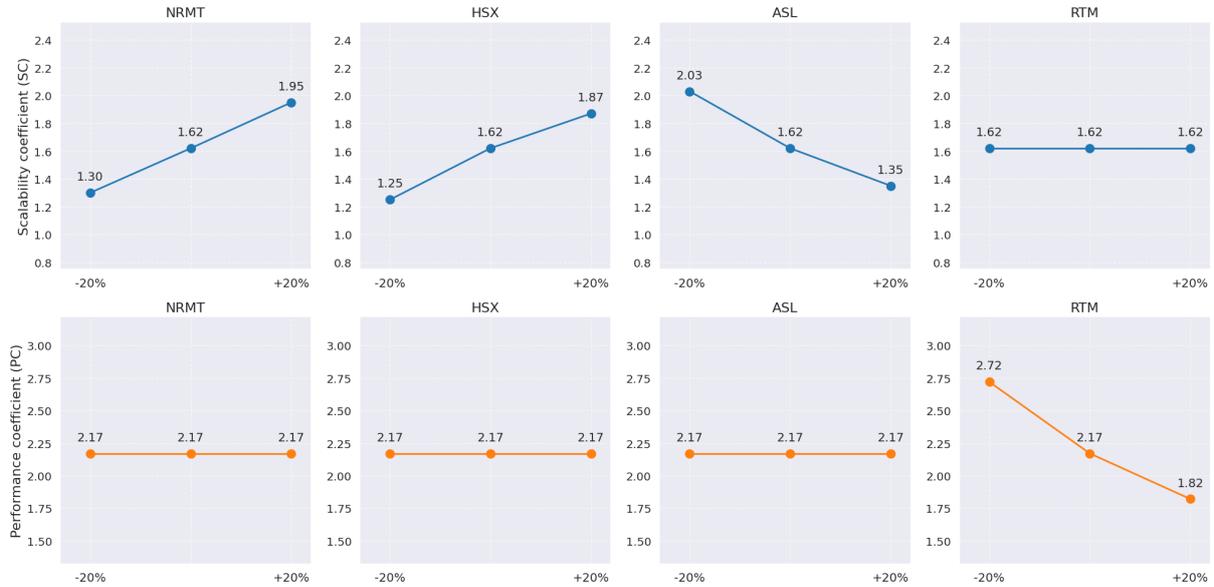

**Fig. 4** *Sensitivity analysis of the Scalability Coefficient (SC) and Performance Coefficient (PC).*

This analysis provides visual understanding about the proposed metrics (*Section 4*) for system architects and developers, guiding their focus towards optimizing the most impactful intrinsic values when designing ML systems for scalability and performance.

## 6.4 Discussions

While this study contributes to the evaluation of architectural patterns in ML-based systems, several limitations and areas for future research should be acknowledged:

A) Metric scope

The proposed evaluation process relies on a limited set of metrics (three each for scalability and performance). Although these provide insights, they may not capture all relevant aspects of ML-based systems. Future work could expand the assessment framework to include additional metrics such as GPU resource utilization, precision, recall, latency, and throughput, offering a more comprehensive evaluation.

B) Microservice implementation

The exact implementation of microservices or localized patterns (*Sections 6.1 and 6.2*) can influence scalability and performance. Factors such as service granularity, communication protocols, and data serialization formats affect system behavior. Future studies should explicitly consider and discuss these aspects.

C) Trade-offs between quality attributes

The current study primarily focuses on scalability and performance. However, improvements in these areas might come at the cost of other quality attributes such as maintainability or reliability. Future research should adopt a more holistic approach, considering these trade-offs and providing guidance on balancing competing quality requirements.

D) Generalizability

The findings' generalizability may be limited due to the focus on a specific ML-based system. To strengthen the validity of the results, the evaluation process should be applied to a diverse set of ML-based systems across different domains and scales. This would help identify common patterns and challenges, as well as provide feedback into the framework's effectiveness in various contexts.

E) Architectural pattern trade-offs

The current study touches on the concept that architectural patterns often present trade-offs, but this aspect deserves more in-depth exploration. In ML systems, a pattern might significantly enhance scalability while introducing a slight performance degradation, or vice versa. For instance, a microservices architecture might improve system scalability but potentially increase latency due to inter-service communication. These nuanced trade-offs are crucial for ML system architects to understand and navigate. Future research should provide a more detailed analysis of these trade-offs, potentially including quantitative measures of the balance between different quality attributes.

F) Evaluation of pattern sets

Another way to strengthen the research could be to provide means for evaluation of a set of patterns instead of individual ones. Extending the existing approach to consider the collective impact of multiple design patterns rather than evaluating them in isolation offers a valuable shift in perspective. Frequently, the implementation of a single pattern may yield improvements in specific areas while inadvertently introducing performance bottlenecks or other issues elsewhere. By examining a suite of patterns as a cohesive unit, we can mitigate isolated variances—referred to as "noise"—that might obscure the overall effectiveness of the patterns when applied in concert.

G) Broader range of ML systems and deployment scenarios

Future work should test the proposed patterns in a broader range of ML systems, not limited to the REST/gRPC and CPU inference scenario. Additionally, it would be beneficial to perform evaluations of the effectiveness of the proposed architectural patterns in more real-world cases.

By addressing these limitations in future work, the robustness and applicability of the proposed assessment framework can be significantly improved, providing even more guidance for the target audience.

## 7. Conclusion

This paper presents two main contributions: (1) a quantitative assessment framework for applying architectural patterns in ML systems, and (2) an evaluation process for these patterns. We focus on scalability and performance, particularly for CPU-based ML inference systems deployed using REST API and gRPC strategies.

Our framework and evaluation process offer a systematic approach to making informed decisions about ML system architecture. By quantitatively assessing the impact of different patterns, software engineers and ML practitioners can make data-driven choices that lead to more efficient and effective systems, especially those dealing with resource-intensive deep neural networks. The proposed evaluation process provides a foundation for further empirical investigations in software design patterns for those systems and addresses unique challenges such as computational complexity, deployment management, and monitoring, which are critical in ML contexts.

Architectural pattern selection should involve close collaboration with stakeholders, considering predefined goals and constraints. Our evaluation process formalizes this, guiding decision-making towards more scalable and high-performant systems.

It's important to recognize that each ML system is unique, and pattern selection should align with specific system requirements. A pattern improving one aspect, like scalability, may introduce new challenges, particularly in CPU-based inference scenarios where resource management is critical.

For future research directions and limitations of this study, we refer to *Section 6.4*. Addressing these areas will further enhance the applicability of this work, contributing to the growing knowledge on software design patterns for ML systems. These efforts will guide the development of more performant, efficient, and scalable ML systems, driving the adoption of ML applications across various domains.

## Acknowledgment

This study is financed by the European Union-NextGenerationEU, through the National Recovery and Resilience Plan of the Republic of Bulgaria, project № BG-RRP-2.004-0008-C01

## Appendix

This appendix provides example code snippets for calculating the metrics used in the quantitative assessment framework, namely *NMRT (Number of Requests at Max Throughput)*, *ASL (Average System Load)*, and *RTM (Response Time Median)*. These code snippets are intended to illustrate the implementation of the metrics and can be adapted to specific use cases.

*Listing 2* demonstrates how to calculate NMRT using Python's requests and *concurrent.futures*. It sends concurrent requests to a specified URL and measures the average response time and total time for each concurrency level. To run this code, make sure to replace "*https://example-target.com/"* with the actual URL of your target system.

```python
import concurrent.futures
import time

import numpy as np
import requests

# Constants
URL = "https://example-target.com/"
START_CONCURRENCY = 10
MAX_CONCURRENCY = 400
STEP = 10
TIMEOUT = 5

def send_request():
    try:
        # Can be adapted to the system endpoint, usually POST
        response = requests.get(URL, timeout=TIMEOUT)
        return response.elapsed.total_seconds()
    except requests.RequestException:
        return None
```

```python
def test_concurrency(concurrency_level):
    with concurrent.futures.ThreadPoolExecutor(
        max_workers=concurrency_level
    ) as executor:
        start_time = time.time()
        futures = [executor.submit(send_request) for _ in range(concurrency_level)]
        results = [f.result() for f in concurrent.futures.as_completed(futures)]
        end_time = time.time()
        results = [r for r in results if r is not None]
        response_time_median = np.median(results) if results else None
        total_time = end_time - start_time
        print(
            f"Concurrency: {concurrency_level}, Response Time Median: {response_time_median}, Total Time: {total_time}"
        )
        return response_time_median, total_time

for concurrency in range(START_CONCURRENCY, MAX_CONCURRENCY + 1, STEP):
    response_time_median, total_test_time = test_concurrency(concurrency)
```

**Listing 2** *Example code for calculating NMRT*

```python
import psutil

load_avg = psutil.getloadavg()
asl = sum(load_avg) / len(load_avg)
print(f"Average System Load (ASL): {asl}")
```

**Listing 3** *Example code for calculating ASL*

```python
import requests
import numpy as np

URL = "https://example-target.com/"
NUM_REQUESTS = 1000

response_times = []

for _ in range(NUM_REQUESTS):
    # Can be adapted to the system endpoint, usually POST
    response = requests.get(URL)
    response_times.append(response.elapsed.total_seconds())

rtm = np.median(response_times)
print(f"Response Time Median (RTM): {rtm} seconds")
```

**Listing 4** *Example code for calculating RTM*